# Exact Enumeration Study of Free Energies of Interacting Polygons and Walks in Two Dimensions

D. Bennett-Wood†, I.G. Enting¶, D. S. Gaunt‡, A. J. Guttmann†*,
J. L. Leask†, A. L. Owczarek† and S. G. Whittington§
†Department of Mathematics and Statistics,
The University of Melbourne, Parkville, Victoria 3052, Australia
¶CSIRO, Division of Atmospheric Research,
Private Bag 1, Mordialloc, Victoria 3195, Australia
‡Department of Physics, King's College,
Strand, London WC2R 2LS, U.K.
§Department of Chemistry,
University of Toronto, Toronto, Canada M5S 3H6

Appeared in *J. Phys. A.*, vol **31**, 4725-4741, (1998).

**Abstract**

We present analyses of substantially extended series for both interacting self-avoiding walks (ISAW) and polygons (ISAP) on the square lattice. We argue that these provide good evidence that the free energies of both linear and ring polymers are equal above the $\theta$-temperature, extending the application of a theorem of Tesi *et. al.* [1] to two dimensions. Below the $\theta$-temperature the conditions of this theorem break down, in contradistinction to three dimensions, but an analysis of the ratio of the partition functions for ISAP and ISAW indicates that the free energies are in fact equal at all temperatures to at least within 1%. Any perceived difference can be interpreted as the difference in the size of corrections-to-scaling in both problems. This may be used to explain the vastly different values of the cross-over exponent previously estimated for ISAP to that predicted theoretically, and numerically confirmed, for ISAW. An analysis of newly extended neighbour-avoiding SAW series is also given.

*email: tonyg@ms.unimelb.edu.au

# 1 Introduction

Long linear polymers in dilute solution are expanded objects under good solvent conditions but if the solvent quality is decreased or, equivalently, the temperature lowered below the $\theta$-temperature, the polymers appear to undergo a sudden collapse transition from an expanded coil to a compact globule. This phenomenon has been studied experimentally by light scattering [2, 3] and by viscosity measurements [4]. In addition, it has been modelled [5] by interacting self-avoiding walks (ISAW) on a lattice with an interaction energy proportional to the number of nearest-neighbour contacts. Considerable progress in the study of this model occurred following the work of de Gennes [6], especially in two dimensions where many critical properties have been determined using Coulomb gas [7] and conformal invariance [8] methods. The model has also been studied numerically using a wide variety of techniques including transfer matrices [9, 10], exact enumeration [11, 12, 13, 14, 15, 16, 17], $1/d$-expansions [18, 19] and Monte Carlo [20, 21, 22, 23, 24, 25, 26, 27, 28, 29]. As a result, for ISAW in $d = 2$ at the $\theta$-point, the critical exponents are believed to take on the values predicted by Duplantier and Saleur [30], in particular the cross-over exponent $\phi = 3/7 = 0.428\ldots$. In addition, the value of the temperature parameter $\beta$ at the collapse transition has been estimated numerically (see eg. [28]) and has a value around $\beta_c = 0.66$ for the square lattice.

A similar collapse transition is believed to occur in randomly branched polymers modelled by lattice animals or lattice trees [31, 32]. For lattice animals, the cross-over exponent and certainly the location of the transition seem to depend on the details of the model. For example, for the $k$-model, a contact model which is a natural generalisation of the ISAW model, $\phi = 0.60 \pm 0.03$, $\beta_c = 0.38 \pm 0.05$ on the square lattice [33], while for the $C$-model, a cycle model, one finds [34] on the same lattice $\phi = 0.657 \pm 0.025$, $\beta_c = 1.87 \pm 0.02$. Both these pairs of values are quite different from the corresponding pair for ISAW quoted above.

On the other hand, there is a growing belief [35, 36, 1, 37] that all models with a given architecture (eg. polygons, uniform $f$-stars, combs, brushes, ... ) not only have the same collapse temperature and cross-over exponent as ISAW but their limiting reduced free energies have the same dependence on the value of the temperature parameter $\beta$ as do ISAW. Let us review the evidence.

First, we define the partition functions for interacting self-avoiding walks (ISAW), polygons (ISAP) and uniform $f$-stars (ISAS-$f$) by

$$Z_n(\beta) = \sum_k c_n(k) e^{\beta k} , \tag{1.1}$$

$$Z_n^o(\beta) = \sum_k p_n(k) e^{\beta k} \tag{1.2}$$

and

$$Z_n(\beta; f) = \sum_k s_n(k; f) e^{\beta k} . \tag{1.3}$$

Here, $c_n(k)$ and $p_n(k)$ are the number of self-avoiding walks and polygons, respectively, with $n$ edges and $k$ contacts, and $s_n(k; f)$ is the number of uniform stars with $f$ branches, $n$ edges in each branch and $k$ contacts. Clearly, $s_n(k; 1) = c_n(k)$. It has been proved rigorously [1, 37] that on a $d$-dimensional simple hypercubic lattice the corresponding limiting reduced free energies

$$\kappa(\beta) = \lim_{n\to\infty} \frac{1}{n} \ln Z_n(\beta) \ , \tag{1.4}$$

$$\kappa^o(\beta) = \lim_{n\to\infty} \frac{1}{n} \ln Z_n^o(\beta) \tag{1.5}$$

and

$$\kappa_f(\beta) = \lim_{n\to\infty} \frac{1}{nf} \ln Z_n(\beta; f) \tag{1.6}$$

exist, and are equal to one another for all $\beta \leq 0$. (More precisely, the proofs by Tesi *et. al.* [1] for walks and polygons are for $d = 3$ but similar arguments should work for general $d$.) Yu *et. al.* [37] also reported, but without proof, that this result extends to uniform combs and brushes.

For $\beta > 0$, the existence of the limiting value $\kappa^o(\beta)$ has been proved rigorously, as has the fact that the limiting function is monotonic and convex [14]. Otherwise, little else has been proved rigorously. However, there is mounting evidence in support of the conjecture

$$\kappa(\beta) = \kappa^o(\beta) = \kappa_f(\beta), \qquad \forall\beta \text{ and } d. \tag{1.7}$$

More specifically, Yu *et. al.* [37] have derived and analysed exact enumeration data for ISAW through orders $n = 25, 18$ and 17, for ISAP through $n = 26, 18$ and 16, for ISAS-3 through $n = 9, 5$ and 6, and for ISAS-4 through $n = 7, 5$ and 4, for the square (SQ), triangular (T) and simple cubic (SC) lattices, respectively. For ISAS-5, the data extend through $n = 4$ (T,SC) and for ISAS-6 through $n = 4$ (T) and 3 (SC). (For $f$-stars, the maximum values of $n$ may seem quite small, but it should be remembered that it is the total number of edges, obtained by multiplying the above values by $f$, that is comparable to the $n$ values for walks and polygons.) The numerical plots of Yu *et. al.* (see Figures 2-4 of [37]) suggest that all these limiting free energies are identical at least up to $\beta = 2$ ($d = 2$) and $\beta = 1.3$ ($d = 3$), both corresponding to temperatures well into the collapsed regions.

Support for $\kappa(\beta) = \kappa_f(\beta)$ for all values of $\beta$ and $d$ comes from their $1/d$-expansions, which Yu *et. al.* [37] derived through order $1/d$ for general $f$, and through order $1/d^2$ for $f = 3$. The terms in the expansions are $\beta$-dependent but turn out to be independent of $f$ and agree term-by-term with the $1/d$-expansion for ISAW, which is known through order $1/d^5$ [19]. However, it has been speculated [36], in the context of the collapse transition for lattice animals, that the range of validity of $1/d$-expansions is limited by the collapse transition at $\beta_c(d)$. If the same happens for ISAW and ISAS-$f$, then the above argument concerning the term-by-term equality of their $1/d$-expansions would have nothing to say when $\beta > \beta_c$.

Support for $\kappa(\beta) = \kappa^o(\beta)$ in $d = 3$ comes from the Monte Carlo results presented by Tesi *et. al.* [1]. They show (see Figure 4 of [1]) that, for the SC lattice, the difference in the relative free energies of ISAW and ISAP, at least up to $\beta = 0.5$ (still well into the collapsed region - see (1.9) below), decreases as $n$ increases (at least up to $n = 1200$), consistent with the limiting free energies being equal for all values of $\beta$. Further confirmation is obtained by making use of theorem 2.8 of [1], which we shall refer to as the *contact theorem*. This proves that if the mean number, $\langle k\rangle_n^o$, of contacts for ISAP is at least as large as the mean number, $\langle k\rangle_n$, for ISAW, at all $\beta > 0$, for $n$ sufficiently large, then the limiting free energies are equal. Tesi *et. al.* studied the behaviour of $\langle k\rangle_n^o/\langle k\rangle_n$ as a function of $\beta$ for several values of $n \leq 1200$ and their Monte Carlo results (see Figures 5 and 6 of [1]) clearly support the equality of the limiting free energies in three dimensions, well into the collapsed region.

We have reviewed the evidence in support of the conjecture (1.7). Assuming now that the conjecture is true implies that the location $\beta_c$ of the collapse transition and the value of the cross-over exponent $\phi$ (using the relation, $\phi = 2 - \alpha$, between $\phi$ and the exponent $\alpha$ characterising the singularity in the free energy at $\beta_c$), are the same for interacting walks, polygons and $f$-stars, as well as, possibly, for other polymer architectures modelled by uniform embeddings of graphs of fixed homeomorphism type. Indeed, there are some direct numerical estimates which are consistent with ISAW and ISAP collapsing at the same value of $\beta$. Thus, in $d = 2$, recent results for ISAW [22, 23, 24, 15, 26, 16, 17, 27, 28] and for ISAP [13, 14] are consistent with a common value around

$$\beta_c = 0.663 \pm 0.016, \qquad \text{(SQ)} \tag{1.8}$$

while in $d = 3$ a common value around

$$\beta_c = 0.277 \pm 0.009, \qquad \text{(SC)} \tag{1.9}$$

is indicated [1, 38].

As for the cross-over exponent $\phi$, there is the conjecture that $\phi = 3/7$ in $d = 2$ [30], while in $d = 3$ — the upper critical dimension for tricritical walks — $\phi$ is believed to take on its mean-field value $\phi = 1/2$ with a leading correction term which is logarithmic [39, 40], for both ISAW and ISAP. As emphasised by Brak *et. al.* [41], cross-over exponents are notoriously difficult to determine numerically. Thus, over the past few years, there has been considerable controversy (see eg. [9, 11, 22, 23, 24, 26]) concerning the value of $\phi$ for ISAW in $d = 2$, with direct numerical estimates ranging from $\phi = 0.48 \pm 0.07$ [9] to $\phi = 0.66 \pm 0.02$ [23]. However, in more recent Monte Carlo work, firstly on the Manhattan lattice, Prellberg and Owczarek [42] found an estimate of $\phi = 0.430 \pm 0.006$ utilising walks of length, $n$, up to $10^6$ and then, with good statistics for $n \leq 2048$, Grassberger and Hegger [27] gave $\phi = 0.435 \pm 0.006$ for the square lattice, both of which seem to confirm the theoretical value of $\phi = 3/7$. Grassberger and Hegger argue that the neglect of extremely large correction-to-scaling terms may have been the cause of the earlier difficulties.

In the case of ISAP in $d = 2$, the best numerical estimate seems to be $\phi = 0.90 \pm 0.02$ [13] which is a long way from $\phi = 3/7$, but is based upon exact enumerations only up to $n = 28$ (that is, 13 terms).

When $d = 3$, most workers (see eg. [1, 38]) have simply accepted the expected theoretical value of $\phi = 1/2$ and we know of no recent direct estimates for either ISAW or ISAP.

In this paper, our aim is to provide additional support for one part of the conjecture (1.7), namely, $\kappa(\beta) = \kappa^o(\beta)$ for $\beta > 0$ and $d = 2$. We do this by first deriving new exact enumeration data for ISAW and ISAP on the square lattice through orders $n = 29$ and $n = 42$, respectively. These new data extend the published data for walks and polygons [43, 13] by 9 and 7 terms and the unpublished data used by Yu *et. al.* [37] by 4 and 8 terms, respectively.

The new data are used, most importantly, for comparing $\langle k \rangle_n$ with $\langle k \rangle_n^o$ for a range of temperatures to determine whether the conditions of the contact theorem [1] are satisfied. This evidence, and that mentioned below, suggest strongly that the free energies of ISAW and ISAP are equal above the $\theta$-temperature ($\beta \leq \beta_c$). Intriguingly however, it seems that the conditions of the contact theorem are *not* satisfied at any temperature below the $\theta$-temperature, using any reasonable extrapolation technique. This coincides with the region where we suggested that the $1/d$-expansions may break down and so are unable to provide an argument for the equality of the ISAW and ISAS-$f$ free energies (the other part of conjecture (1.7)). However, we have directly estimated the difference in the limiting free energies of ISAP and ISAW,

$$\Delta\kappa \equiv \kappa^o(\beta) - \kappa(\beta) \; , \tag{1.10}$$

as a function of $\beta$, and found that for a wide range deep into the collapsed phase this difference is $0.00 \pm 0.01$ (where $\kappa$ and $\kappa^o$ are of the order 1.0). At most high temperatures the error is considerably smaller (about 0.001). We have supplemented the exact enumeration data by simulating ISAP using a Monte Carlo algorithm which we argue only provides reliable information for $n$ well below 1000 at the temperatures required, at, and below, the collapse temperature. The analysis of the data illustrates the near impossibility of extracting reliable direct estimates of the critical parameters from data of this order of $n$. Hence we propose that radically new algorithms are needed to simulate ISAP near, and especially below, the $\theta$-temperature in $d = 2$. Umbrella sampling and multiple Markov chain methods (see [29] and references therein) have proved to be successful in $d = 3$ and these are promising techniques for future work in $d = 2$.

The paper is divided as follows. In section 2 we describe the exact enumeration and Monte Carlo techniques utilised. In section 3 we present our analyses and discuss their meaning. We conclude with a short summary of our results.

## 2 Data Derivation

In this section, we describe the methods that we have used to extend the exact enumeration data for ISAP and ISAW on the square lattice, and the details of the Monte Carlo algorithm used to simulate ISAP.

### 2.1 Finite Lattice Method

Exact enumeration results, giving the complete polynomials in $w = e^{\beta}$, were obtained for all square lattice polygons with perimeter up to $n = 28$ by Maes and Vanderzande [13]. We have used the finite lattice method to extend these data by seven terms up to $n = 42$. (Only terms with perimeters of even length contribute, of course.) We take the opportunity to correct a small error in Table I of [13]; the number $C(26, 9)$ should be 679 848, rather than 679 484 as printed.

The finite lattice method of enumerating self-avoiding polygons (SAP) on the square lattice was first introduced by Enting [44] — the enumeration extending to polygons with $n = 38$, increasing the number of terms known at that time by over 50%. Later work extended this enumeration to $n = 56$ [45, 46], and currently stands at $n = 70$ [47]. It has also been possible to augment these enumerations with calculations of other geometrical properties of the polygons. Thus, calliper moments up to $n = 54$ were obtained by Guttmann and Enting [46] and the enumeration of polygons by both perimeter and area was given by Enting and Guttmann [48].

The technique has also been applied to other planar lattices. Series for the L and Manhattan lattices were obtained up to $n = 48$ [45] and recently extended to $n = 84$ [49], for the honeycomb lattice up to $n = 82$ [50], and for the triangular lattice up to $n = 25$ [51] classified by both perimeter and area.

The quantity that we wish to determine is $p_n(k)$ in (1.2), the number of square lattice unrooted polygons with $n$ edges and $k$ nearest-neighbour contacts. For convenience in this section, we write $p_{n,k} \equiv p_n(k)$, so that (1.2) becomes

$$Z_n^o(w) = \sum_k p_{n,k} w^k \ . \tag{2.1}$$

For any fixed $n$, the $p_{n,k}$ are all zero for $k > k_{max}(n)$ since the sum on the right-hand side of (2.1) is a polynomial in $w$. Since each site of the polygon can be involved in at most two near-neighbour contacts and since each near-neighbour contact involves two sites, we must have $k_{max}(n) \leq n$. In practice, we can set $k_{max}$ empirically and use the total $n$-edge polygon count to ensure that we have used a sufficiently large value. Finally, we define the generating function

$$C(x, w) = \sum_n Z_n^o(w)\, x^n = \sum_{k,n} p_{n,k}\, w^k\, x^n \ . \tag{2.2}$$

The finite lattice method of enumerating polygons involves two steps. Firstly, we need to enumerate polygons constrained to lie within various finite rectangles. The second step is that such enumerations for finite rectangles are combined to give a truncated approximation of the infinite lattice polygon generating function. If the factors used in the linear combination of the finite lattice generating functions are chosen correctly, then the first incorrect term in the infinite lattice generating function will correspond to the largest polygon that cannot be embedded in any of the rectangles that are used.

When enumerating polygons constrained to lie within various finite rectangles one needs to classify the polygons according to some quantity such as perimeter or area that grows as the size of the rectangles increases. Such a quantity forms the expansion variable of the generating function. There is, however, no need to confine the classification to just one quantity; apart from the (non-trivial) overhead of working with series in two variables, the finite lattice method applies to polygon enumerations involving several variables, as (for example) in the enumeration of polygons by perimeter and area [50, 48].

The weights $a_{\ell,m}$ used to combine the finite lattice generating functions are used in the expression

$$C(x,w) \approx \sum_{\ell,m} a_{\ell,m}\, G_{\ell,m}\ , \tag{2.3}$$

where $G_{\ell,m}$ is the generating function for polygons that can be embedded in a rectangle of width $\ell$ and length $m$ so as to span the length of the rectangle. The lattice symmetry gives some degree of choice in the weights. Enting and Guttmann (see eqn. 2.8a-e in [45]) give the weights used in most of the polygon enumerations. However, in this calculation, we use a slightly different formulation given by Guttmann and Enting [46] when enumerating calliper moments, even though we do not retain the calliper moments in the present calculation.

Let $p_{n,k;q}$ be the number of $n$ edge polygons with $k$ contacts which span a distance $q$ in the horizontal direction, i.e. the difference between the maximum and minimum values of the horizontal coordinates is $q$ lattice units. Then, the $j^{\text{th}}$ calliper moment is

$$C^{[j]}(x,w) = \sum_{n,k,q} q^j p_{n,k;q}\, w^k\, x^n\ . \tag{2.4}$$

We define

$$S_q(x,w) = \sum_{n,k} p_{n,k;q}\, w^k\, x^n\ , \tag{2.5}$$

whence

$$C^{[j]}(x,w) = \sum_{q} q^j S_q(x,w)\ . \tag{2.6}$$

The finite lattice approximations for the $S_q$ are

$$S_q(x,w) = \sum_{j=1}^{2N+1-q} \left(G_{q,j} - 2G_{q-1,j} + G_{q-2,j}\right)\ , \qquad q \le N\ , \tag{2.7}$$

and

$$S_q(x,w) = G_{2N+1-q,q} - G_{2N-q,q} \; . \tag{2.8}$$

(This last expression also corrects a minor typographical error in eqn. (6) of Guttmann and Enting [46].) These expressions give the correct enumeration of polygons of up to $4N+2$ edges (and their calliper moments).

The enumeration of the generating functions, $G_{\ell,m}$, for polygons in rectangles uses a transfer matrix technique. The basic technique is as described by Enting [44] with the formalism extended to take account of nearest-neighbour contacts. The transfer matrix technique works with the generating functions for sets of loops in partly constructed lattices (shown by solid points on Figure 1). The boundary of the partly constructed lattice is defined by a transect line drawn on the dual lattice, as shown by the dashed line in Figure 1. Each step of the construction involves moving the transect line (from the dashed position to the dotted position) so as to add one site (shown as circled) and two new edges (shown as +++++) to the partly completed lattice. The construction of self-avoiding polygons only requires a knowledge of how the bonds of the polygon intersect the transect line. As described by Enting [44], it is sufficient to distinguish between edges with no bond of the polygon (denoted '0'), edges with a bond of the polygon that is the uppermost arm of a loop (denoted '1') and edges with a bond that is the lower arm of a loop (denoted '2'). This '1','2' notation uniquely specifies the connectivity of the loops of the partly constructed polygon. For the present study, we need to add a new edge state ('3') to denote edges along which a nearest-neighbour contact occurs or, more precisely, edges along which a nearest-neighbour contact may occur if an occupied site is added at the end of a type '3' bond, as the lattice is constructed.

The process of adding a site (shown as double circle in Figure 1) consists of linking the two incoming edges (shown with double lines in Figure 1) that occur at the kink in the transect line and assigning states to the two new edges (shown as +++++ in Figure 1) leaving the site. Enting [44] lists the allowed outputs for all allowed combinations of '0', '1', '2' on the input edges, where '0' corresponds to no step, '1' and '2' to the uppermost (lowermost) end of a loop, respectively. The rules for the present case are the same with the following re-interpretations. For the purposes of classifying input states, type '3' is equivalent to type '0'. An output state of '0' in the tabulation of Enting [44] is replaced by '3' except for the case when a $(0,0)$ pair is transformed to a new $(0,0)$ pair. A factor of $w^{n(3)}$, where $n(3)$ is the number of incoming type '3' edges, is included except when the output is $(0,0)$. A summary of these rules is given in Table 1.

The use of '1' and '2' to denote upper and lower ends of loops constrains the relative arrangements of such edges, but the '0' and '3' can be interspersed freely among them. (The same situation applies on the triangular lattice where we classify sites according to 4 states.) The number of configurations needed to enumerate polygons with nearest-neighbour contacts on a rectangle of width $W$ is the same as the number of configurations used when enumerating triangular lattice polygons on a strip of width $W+1$.

The maximum width that we have used is 10 and so we have been able to enumerate polygons of up to 42 steps, with a complete specification of the distribution of nearest-neighbour contacts. This limit is imposed by storage requirements rather than time limitations. Our results are given in Appendix A.

### 2.2 Direct enumeration

The direct enumeration of ISAW on the square lattice, giving the complete polynomials in $w$, were given by Ishinabe [43] through $n = 20$ (see Table 1 of [43]). Five more terms were derived, but not published, by Yu *et. al.* (1997) in their study of the free energies of ISAW, ISAP and ISAS-$f$. We have used a 52 processor Intel Paragon supercomputer to extend the direct enumeration of ISAW up to $n = 29$. All the terms, $Z_1$ through $Z_{29}$, are given in Appendix B. The calculation took about 100 hours.

Recently, a 1024 processor Intel Paragon supercomputer was used by Conway and Guttmann [52] to implement a finite lattice method which extended the enumeration of square lattice self-avoiding walks (SAW) from $n = 39$ to $n = 51$. Unfortunately, the parallelised algorithm, which is challenging to implement efficiently, has not so far been generalised to enumerate ISAW.

As the temperature parameter $\beta$ approaches $-\infty$, the walks become neighbour avoiding. In this special case we have extended the series to 32 terms. The coefficients are also given in Appendix B, as they are the coefficients $c_n(0)$.

### 2.3 Monte Carlo Algorithm

Monte Carlo estimates of thermodynamic properties, including the energy and heat capacity, were obtained at a series of temperatures and lengths of polygons. The algorithm was a basic Metropolis algorithm involving sampling along a realisation of a Markov chain whose unique limit distribution was the Boltzmann distribution at the required temperature. The underlying symmetric Markov chain was first defined for walks by a set of pivot moves [53, 54] combined with local moves to improve the “slow mode” problem associated with near-neighbour contacts. For polygons we hence used the corresponding “cut-and-paste” algorithm invented by Madras *et. al.* [55]. This algorithm works well, in the sense that the autocorrelation times of the various observables are short, for high temperatures, but less well at lower temperatures. However, for values of $n$ less than 200 we were able to sample effectively at temperatures down to just below that corresponding to the maximum in the heat capacity, so that the polygon was just inside the collapsed regime.

## 3 Free Energy and Contact Number Analysis

The exact enumeration data given in Appendices A and B for ISAP and ISAW respectively, were firstly used to calculate the expected number of contacts for each model at a range of values of $\beta$. For $\beta \leq 0.663$, we plotted the contact densities for ISAW and ISAP, $m_n(\beta) = \langle k \rangle_n / n$ and $m^o_n(\beta) = \langle k \rangle^o_n / n$, respectively, against $1/n$ on the same graph, while for $\beta > 0.663$ we plotted these against $1/\sqrt{n}$ — six of these plots are given in Figure 2. We have used these scales since these corrections are expected from the partition function scaling forms most likely in each regime [56].

For small $\beta < 0.6$ the polygon data, $m^o_n$, are larger than the walk data, $m_n$, at the largest values of $n$: there is a crossing point at low $n$ which moves to larger $n$ as $\beta$ increases. Using linear and quadratic fits, and adding in $n^{-3/2}$ corrections allowed us to estimate the thermodynamic limit for $m_n(\beta)$ and $m^o_n(\beta)$ which we denote $m(\beta)$ and $m^o(\beta)$ respectively. If the thermodynamic limit free energies are the same so should these limiting contact densities, and for $\beta < 0.663$ this is true within 0.5%. It is also true, that if anything, the extrapolations would infer that $m^o \geq m$, but the contact theorem then implies that the free energies and hence these average contact densities are equal. Near and above the estimated $\theta$-temperature our extrapolations lead us to believe that $m^o_n(\beta) \geq m_n(\beta)$ for $n$ large enough (that is, there is at least a crossover point beyond the extent of the series), which would again allow the use of the contact theorem. Hence we deduce that $m(\beta) = m^o(\beta)$ for all $\beta < 0.663$

For temperatures below the critical temperature ($\beta \geq 0.663$), the crossing point of the walk and polygon data disappears (see Figure 2). It is unclear whether there may be one at large $n$ or whether the $\theta$-temperature marks a point where the crossing point moves off to $\infty$, thereby marking the beginning of a regime where, for all $n$, $m_n > m^o_n$. The conditions of the contact theorem would no longer apply, admitting the *possibility* that the ISAW and ISAP free energies may be different. In addition, our extrapolations were now far more sensitive to the number of terms used, and on small variations of the extrapolation function, especially for the walk data. The values of $m(\beta)$ and $m^o(\beta)$ may reasonably differ by up to 5% if any fairly conservative extrapolation is taken seriously.

For finite $n$, the curves of the contact densities are substantially different for ISAW and ISAP, and further differentiation, giving specific heats and third cumulants (the contact density curve is the first derivative of the free energy in the variable $\beta$), produce radically different graphs. Moreover, extrapolations seem to need different extrapolation functions to obtain results consistent with the equality of the free energies at all $\beta$. It is no wonder that past use of the ISAP data has produced vastly different results to the ISAW data. We have also previously simulated self-avoiding polygons with a cut-and-join algorithm (see subsection (2.3)) essentially in an attempt to find the $\theta$-temperature and cross-over exponent for ISAP, without great success. While we were able to simulate lengths up to 1000, long auto-correlation times restricted the use of the data to around maximum lengths of 100 to 200. The size of the corrections-to-scaling in ISAP and their clear difference in magnitude, manifested in the contact data described above, rendered that analysis less useful than we had hoped. We do point out that using the third cumulant, which is expected to diverge at $\beta_c$ we estimated $\phi$ to be around 0.5. (Although had we chosen higher derivatives, different values would have ensued.) We do not give error bars since we do not believe that convergence has been achieved. While this cannot be used to confirm the theoretical value of 3/7 for ISAP, it does throw considerable doubt on the value of $0.90 \pm 0.02$, previously quoted [13]. We remark that to reconcile the ISAW and ISAP results near and below the $\theta$-temperature with Monte Carlo data will require simulation of ISAP of lengths over 1000 with good statistics: this is something the cut-and-join algorithm and current computing power seem unable to achieve. However, we were able to use our Monte Carlo data to reinforce our conclusions concerning the extrapolations of the expected density of contacts mentioned above, in each regime.

To clarify the low temperature situation and provide additional support for the conclusions at higher temperatures, we have analysed the series formed by the ratio of the polygon partition function to the walk partition function over the same range of temperature considered above. The standard method of differential approximants was used [57], with a statistical averaging procedure over a wide range of inhomogeneous approximants applied. To use even and odd $n$, we in fact analysed the series $Q_n(\beta)$, where

$$Q_n = \begin{cases} \frac{\sqrt{Z^o_{n+1} Z^o_{n-1}}}{Z_n} & \text{for } n \text{ odd} \\ \frac{Z^o_n}{Z_n} & \text{for } n \text{ even.} \end{cases} \tag{3.1}$$

If the expected asymptotic behaviour occurs for the partition functions (including the equality of the ISAW and ISAP free energies), the quantity $Q_n$ [16] should behave as a power law with connective constant 1: the exponent of the power law will be different below, at, and above the $\theta$-temperature. Here we are not interested in identifying this power law or the value of its exponent, only in verifying that the connective constant is indeed 1, since it is this fact that implies the equality of the ISAW and ISAP free energies. The difference in free energy is given by

$$\Delta\kappa = \ln \left( \lim_{n\to\infty} (Q_n)^{1/n} \right) . \tag{3.2}$$

We examined a range of $\beta$ from 0 to 1.5 and the value of $\Delta\kappa$ was 0.0000 within the errors found. The errors for the following $\beta$, that is 0, 0.2, 0.4, 0.6, 0.663, 0.8, 1.0, 1.2, and 1.5, were 0.0004, 0.0004, 0.0006, 0.002, 0.002, 0.002, 0.006, 0.008, and 0.01, respectively. These data would seem to imply that the conjecture (1.7) holds for ISAW and ISAP at all temperatures despite the conditions of the contact theorem probably failing and the possible breakdown in the $1/d$-expansions.

While several authors in the past have plotted $\kappa(\beta)$ for ISAW, we give below a table of values of $\kappa^o(\beta)$ found from a differential approximant analysis of the $Z^o_n$ series: see Table 2. We note that the values for $\beta < 1.0$ fall within graphical accuracy on the curve drawn by Nidras [28] following his analysis of Monte Carlo data for ISAW.

For $\beta = -\infty$ we have neighbouring-avoiding walks and polygons. This is an interesting problem in its own right. We have analysed the series given in Appendix B by the standard method of differential approximants and find for walks

$$\text{unbiased} \quad x_c = 0.43180(2)\,, \qquad \gamma = 1.344(1) \tag{3.3}$$

$$\text{biased} \quad x_c = 0.4317925(1) \tag{3.4}$$

where the biased value has imposed $\gamma = 43/32$ on the approximants, and $x_c$ (from which the free energy can be calculated) is the closest singularity to the origin of the generating function of partition functions. For polygons (noting that $x_c$ for polygons should be equal to the square of $x_c$ for walks since only even length polygons exist) we obtain

$$\text{unbiased} \quad x_c = 0.1867(6)\,, \qquad 2-\alpha = 1.5(2) \tag{3.5}$$

$$\text{biased} \quad 2-\alpha = 1.43(16)\ \text{(first order)}, \quad 1.47(11)\ \text{(second order)} \tag{3.6}$$

where the biased exponent estimates have been obtained using the value given in (3.4) for the critical point. The exponent estimates are consistent with the expected value of $\alpha = 1/2$.

# 4 Summary

We have presented and analysed substantially extended series for both interacting self-avoiding walks (ISAW) and polygons (ISAP) on the square lattice. Our analysis provides good evidence that the free energies of both linear and ring polymers are equal above the $\theta$-temperature; this result is consistent with an extension of a theorem of Tesi *et. al.* [1], but below the $\theta$-temperature the conditions of this theorem break down. However, an analysis of the ratio of the partition functions for ISAP and ISAW indicate that the free energies are in fact equal at all temperatures to at least within 1%. Any perceived difference can be interpreted as the difference in the size of corrections-to-scaling in both problems. This may explain the vastly different values of the cross-over exponent previously estimated for ISAP to that predicted theoretically, and numerically confirmed, for ISAW. We also present newly extended neighbour-avoiding SAW series and analyse them. We develop a Monte Carlo approach to this problem, and discuss its application to ISAPs.

# Acknowledgements

The authors take pleasure in thanking R. Brak for carefully reading the manuscript. Financial support from the Australian Research Council is gratefully acknowledged by AJG, JLL, and ALO. Some of this work was carried out while SGW and DSG were separately visiting the Department of Mathematics and Statistics of the University of Melbourne, and they are grateful to the members of that department for their kind and generous hospitality.

## A Interacting Self-Avoiding Polygon Enumerations

We given below in Table 3 the coefficients $p_n(k)$ for the numbers of self-avoiding polygons of length $n$ with $k$ nearest-neighbour contacts up to $n = 42$.

## B Interacting Self-Avoiding Walk Enumerations

We given below in Table 4 the coefficients $c_n(k)$ (actually we give $c_n(k)/4$) the numbers of self-avoiding walks of length $n$ with $k$ nearest-neighbour contacts up to $n = 29$. When $k = 0$ neighbour-avoiding walks are realised and the numbers, $c_n(0)$, in Table 4 give the numbers of neighbour-avoiding walks up to $n = 32$.

## Figure Captions

Figure 1. The way in which a transect line (dashed) is drawn through the square lattice, cutting $W + 2$ edges for a rectangle of width $W$. The dotted line shows the new position of the transect line after the elementary step of adding one new site (circled) and two new outgoing edges +++++ to replace two old incoming edges (double).

Figure 2 The six graphs are plots of the expected number of contacts $m_n(\beta)$ and $m_n^o(\beta)$ for ISAW (circles with crosses) and ISAP (crosses) respectively at six different (fixed) values of $\beta$. For $\beta = 0.2$, 0.4, and 0.6, which are expected to lie in the expanded phase we have plotted the two sequences $m_n$ against $1/n$, while for $\beta = 0.8$, 1.0, and 1.5, which are expected to lie in the collapsed regime, we have used $1/\sqrt{n}$. These scales were chosen to reflect the expected corrections-to-scaling in those regimes, which in turn reflect the expected asymptotic forms of the partition function scaling.

## Table Captions

Table 1. Rules for allowed states of outgoing edges (x,y) for all possible states of incoming edges. The new partial generating function incorporates a factor of $x^{n(1)+n(2)}$, where $n(j)$ is the number of outgoing edges of type $j$, and a factor of $w^{k(3)}$, where $k(3)$ is the number of incoming edges of type '3', except in the case '*' where no bonds pass through the site. In the cases marked †, other edges must be relabelled as specified by Enting [44]. In the case marked ‡ there is no new state, but the partial generating function is included in the running total for $C(x,w)$ with the appropriate $a_{\ell,m}$ factor.

Table 2. A list of estimates of the reduced free energy of ISAP from a differential approximant analysis. The value for $\beta = -\infty$ is obtained from walk data.

Table 3. The coefficients $p_n(k)$ for $n \leq 42$.

Table 4. The coefficients $c_n(k)/4$ for $n \leq 29$ and all $k$, and for $k = 0$ with $n \leq 32$.

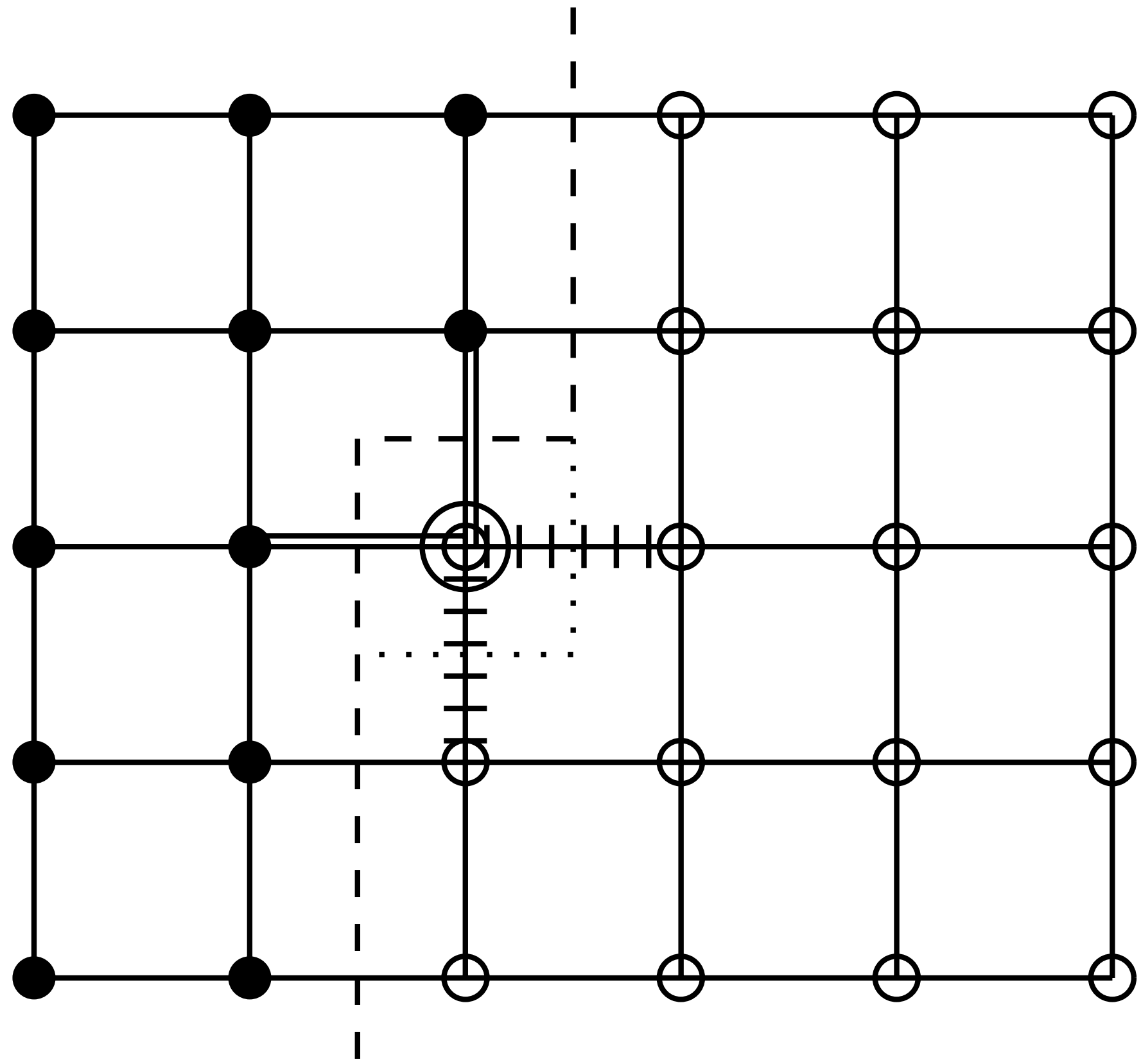

Figure 1:

**Title:** Exact Enumeration Study of Free Energies of Interacting Polygons and Walks in Two Dimensions

**Authors:** D. Bennett-Wood *et. al.*

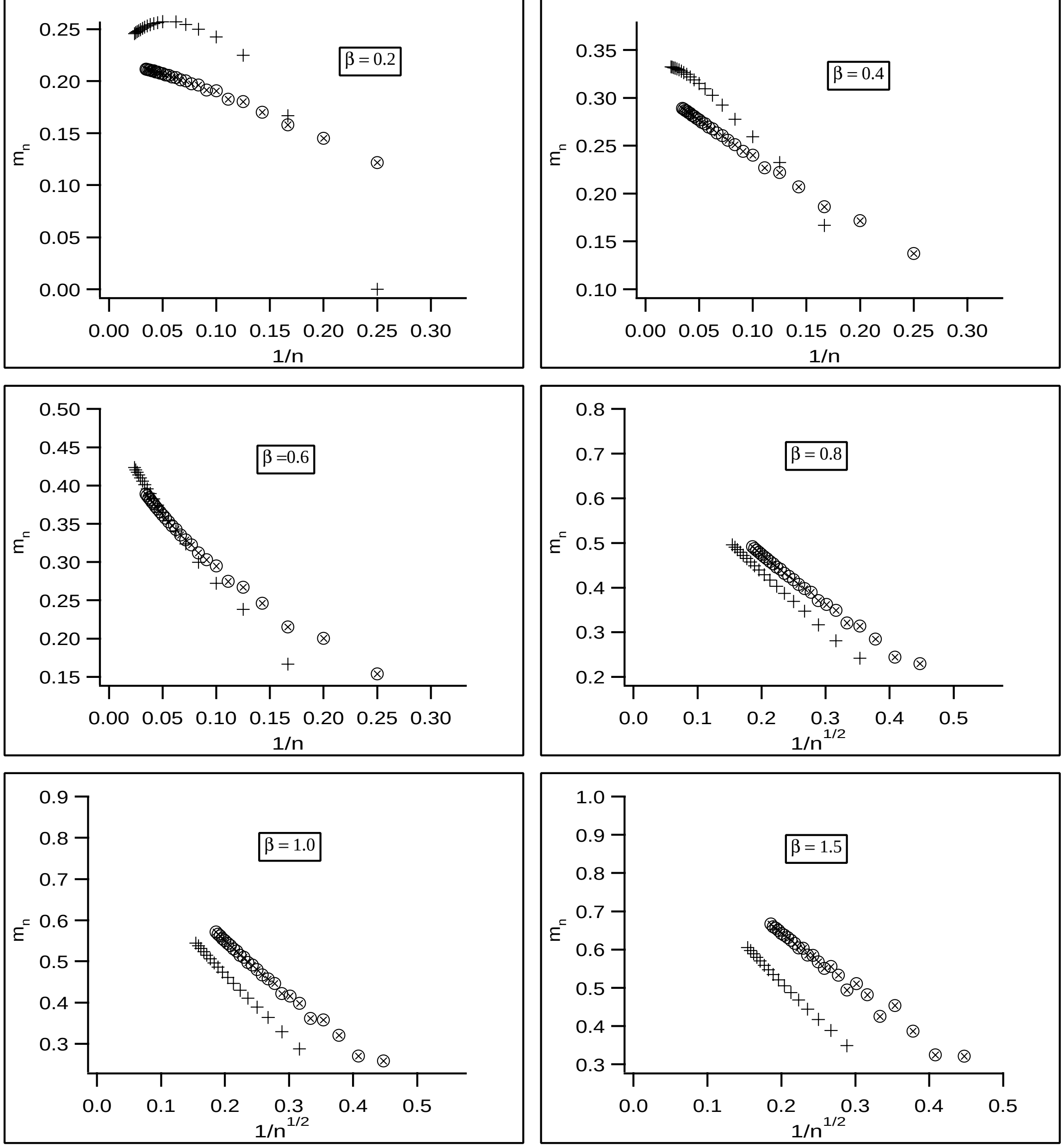

Figure 2:

**Title:** Exact Enumeration Study of Free Energies of Interacting Polygons and Walks in Two Dimensions

**Authors:** D. Bennett-Wood *et. al.*

## Tables

| Inputs: | (0,1)<br>(3,1)<br>(1,0)<br>(3,0) | (0,2)<br>(3,2)<br>(2,0)<br>(2,3) | (0,0)<br>(0,3)<br>(3,0)<br>(3,3) | (1,2) | (2,1) | (1,1)<br>(2,2) |
|---|---|---|---|---|---|---|
| Outputs: | (1,3)<br>(3,1) | (2,3)<br>(3,2) | (0,0)*<br>(1,2) | ‡ | (3,3) | (3,3)† |

Table 1:

| $\beta$ | $\kappa^o$ |
|---|---|
| $-\infty$ | 0.839810(7) |
| $-2.0$ | 0.8542(3) |
| $-1.0$ | 0.8816(2) |
| $-0.5$ | 0.91194(6) |
| $-0.25$ | 0.936344(3) |
| 0 | 0.9700811(1) |
| 0.2 | 1.007(1) |
| 0.4 | 1.060(2) |
| 0.6 | 1.141(5) |
| 0.663 | 1.170(4) |
| 0.8 | 1.254(5) |
| 1.0 | 1.40(2) |
| 1.2 | 1.55(3) |
| 1.5 | 1.79(5) |

Table 2:

| $n$ | $k$ | $p_n(k)$ | $n$ | $k$ | $p_n(k)$ | $n$ | $k$ | $p_n(k)$ | $n$ | $k$ | $p_n(k)$ |
|---|---|---|---|---|---|---|---|---|---|---|---|
| 4 | 0 | 1 | 6 | 0 | 0 | 8 | 0 | 1 | 10 | 0 | 2 |
| | | | 6 | 1 | 2 | 8 | 1 | 0 | 10 | 1 | 8 |
| | | | | | | 8 | 2 | 6 | 10 | 2 | 0 |
| | | | | | | | | | 10 | 3 | 18 |
| 12 | 0 | 9 | 14 | 0 | 36 | 16 | 0 | 154 | 18 | 0 | 668 |
| 12 | 1 | 20 | 14 | 1 | 96 | 16 | 1 | 408 | 18 | 1 | 1832 |
| 12 | 2 | 40 | 14 | 2 | 110 | 16 | 2 | 562 | 18 | 2 | 2564 |
| 12 | 3 | 0 | 14 | 3 | 156 | 16 | 3 | 488 | 18 | 3 | 2704 |
| 12 | 4 | 51 | 14 | 4 | 16 | 16 | 4 | 584 | 18 | 4 | 2218 |
| 12 | 5 | 4 | 14 | 5 | 138 | 16 | 5 | 176 | 18 | 5 | 2292 |
| | | | 14 | 6 | 36 | 16 | 6 | 372 | 18 | 6 | 1074 |
| | | | | | | 16 | 7 | 188 | 18 | 7 | 1076 |
| | | | | | | 16 | 8 | 6 | 18 | 8 | 740 |
| | | | | | | | | | 18 | 9 | 100 |
| 20 | 0 | 2932 | 22 | 0 | 13016 | 24 | 0 | 58364 | 26 | 0 | 264208 |
| 20 | 1 | 8372 | 22 | 1 | 38876 | 24 | 1 | 183044 | 26 | 1 | 871596 |
| 20 | 2 | 12388 | 22 | 2 | 60918 | 24 | 2 | 304010 | 26 | 2 | 1533190 |
| 20 | 3 | 13464 | 22 | 3 | 70350 | 24 | 3 | 370780 | 26 | 3 | 1971494 |
| 20 | 4 | 12983 | 22 | 4 | 69208 | 24 | 4 | 382224 | 26 | 4 | 2118120 |
| 20 | 5 | 10368 | 22 | 5 | 62212 | 24 | 5 | 348888 | 26 | 5 | 2010196 |
| 20 | 6 | 9194 | 22 | 6 | 47482 | 24 | 6 | 292470 | 26 | 6 | 1718270 |
| 20 | 7 | 5120 | 22 | 7 | 37628 | 24 | 7 | 214628 | 26 | 7 | 1360788 |
| 20 | 8 | 3679 | 22 | 8 | 22364 | 24 | 8 | 158126 | 26 | 8 | 969218 |
| 20 | 9 | 2532 | 22 | 9 | 14490 | 24 | 9 | 95828 | 26 | 9 | 679848 |
| 20 | 10 | 766 | 22 | 10 | 8604 | 24 | 10 | 59986 | 26 | 10 | 414052 |
| 20 | 11 | 28 | 22 | 11 | 3924 | 24 | 11 | 32256 | 26 | 11 | 250622 |
| | | | 22 | 12 | 500 | 24 | 12 | 16232 | 26 | 12 | 132908 |
| | | | | | | 24 | 13 | 4280 | 26 | 13 | 63386 |
| | | | | | | 24 | 14 | 154 | 26 | 14 | 24452 |
| | | | | | | | | | 26 | 15 | 3028 |

Table 3:

| $n$ | $k$ | $p_n(k)$ | $n$ | $k$ | $p_n(k)$ | $n$ | $k$ | $p_n(k)$ | $n$ | $k$ | $p_n(k)$ |
|---|---|---|---|---|---|---|---|---|---|---|---|
| 28 | 0 | 1206818 | 30 | 0 | 5558724 | 32 | 0 | 25803509 | 34 | 0 | 120638466 |
| 28 | 1 | 4189420 | 30 | 1 | 20297228 | 32 | 1 | 99008272 | 34 | 1 | 485808492 |
| 28 | 2 | 7791274 | 30 | 2 | 39822158 | 32 | 2 | 204447542 | 34 | 2 | 1053436400 |
| 28 | 3 | 10541380 | 30 | 3 | 56574708 | 32 | 3 | 304436224 | 34 | 3 | 1641203412 |
| 28 | 4 | 11805811 | 30 | 4 | 66024666 | 32 | 4 | 369974212 | 34 | 4 | 2075439970 |
| 28 | 5 | 11601068 | 30 | 5 | 67216160 | 32 | 5 | 390203512 | 34 | 5 | 2266884096 |
| 28 | 6 | 10285214 | 30 | 6 | 61558578 | 32 | 6 | 369111558 | 34 | 6 | 2214114652 |
| 28 | 7 | 8337688 | 30 | 7 | 51656214 | 32 | 7 | 319477936 | 34 | 7 | 1975494948 |
| 28 | 8 | 6320269 | 30 | 8 | 40178374 | 32 | 8 | 256686755 | 34 | 8 | 1634546818 |
| 28 | 9 | 4399656 | 30 | 9 | 29443298 | 32 | 9 | 193161096 | 34 | 9 | 1267837116 |
| 28 | 10 | 2975016 | 30 | 10 | 20083644 | 32 | 10 | 137613088 | 34 | 10 | 927667754 |
| 28 | 11 | 1808576 | 30 | 11 | 13178456 | 32 | 11 | 92079812 | 34 | 11 | 645059158 |
| 28 | 12 | 1057622 | 30 | 12 | 7968438 | 32 | 12 | 59007648 | 34 | 12 | 424295022 |
| 28 | 13 | 567540 | 30 | 13 | 4551574 | 32 | 13 | 35428684 | 34 | 13 | 266938184 |
| 28 | 14 | 262116 | 30 | 14 | 2446186 | 32 | 14 | 19977836 | 34 | 14 | 158957976 |
| 28 | 15 | 112192 | 30 | 15 | 1153074 | 32 | 15 | 10655808 | 34 | 15 | 89006190 |
| 28 | 16 | 27560 | 30 | 16 | 483900 | 32 | 16 | 5163928 | 34 | 16 | 47110136 |
| 28 | 17 | 1204 | 30 | 17 | 166728 | 32 | 17 | 2163628 | 34 | 17 | 23154978 |
| | | | 30 | 18 | 22112 | 32 | 18 | 818630 | 34 | 18 | 10030816 |
| | | | 30 | 19 | 308 | 32 | 19 | 199836 | 34 | 19 | 3814080 |
| | | | | | | 32 | 20 | 13146 | 34 | 20 | 1238968 |
| | | | | | | | | | 34 | 21 | 191868 |
| | | | | | | | | | 34 | 22 | 4864 |

Table 3: continued

| $n$ | $k$ | $p_n(k)$ | $n$ | $k$ | $p_n(k)$ | $n$ | $k$ | $p_n(k)$ |
|---|---|---|---|---|---|---|---|---|
| 36 | 0 | 567732133 | 38 | 0 | 2687937916 | 40 | 0 | 12796823923 |
| 36 | 1 | 2396065580 | 38 | 1 | 11871631876 | 40 | 1 | 59058603772 |
| 36 | 2 | 5444273148 | 38 | 2 | 28208568050 | 40 | 2 | 146480771246 |
| 36 | 3 | 8859088348 | 38 | 3 | 47864522384 | 40 | 3 | 258774823792 |
| 36 | 4 | 11647025630 | 38 | 4 | 65356204120 | 40 | 4 | 366598059998 |
| 36 | 5 | 13169303200 | 38 | 5 | 76458854924 | 40 | 5 | 443457246668 |
| 36 | 6 | 13275125086 | 38 | 6 | 79511992592 | 40 | 6 | 475535061978 |
| 36 | 7 | 12202175232 | 38 | 7 | 75247286342 | 40 | 7 | 463109497364 |
| 36 | 8 | 10388601811 | 38 | 8 | 65868482586 | 40 | 8 | 416535589099 |
| 36 | 9 | 8284253876 | 38 | 9 | 53955153548 | 40 | 9 | 350206578292 |
| 36 | 10 | 6237262394 | 38 | 10 | 41706504166 | 40 | 10 | 277676883848 |
| 36 | 11 | 4454547472 | 38 | 11 | 30609613076 | 40 | 11 | 208991818212 |
| 36 | 12 | 3034098341 | 38 | 12 | 21408550092 | 40 | 12 | 150036552328 |
| 36 | 13 | 1965904908 | 38 | 13 | 14324532228 | 40 | 13 | 103039789644 |
| 36 | 14 | 1218682494 | 38 | 14 | 9157864514 | 40 | 14 | 67883242114 |
| 36 | 15 | 719371560 | 38 | 15 | 5608439482 | 40 | 15 | 42879298560 |
| 36 | 16 | 401071891 | 38 | 16 | 3282544430 | 40 | 16 | 25992755022 |
| 36 | 17 | 211258692 | 38 | 17 | 1824153318 | 40 | 17 | 15097553036 |
| 36 | 18 | 104443870 | 38 | 18 | 958806512 | 40 | 18 | 8363957186 |
| 36 | 19 | 46753216 | 38 | 19 | 475774598 | 40 | 19 | 4395853816 |
| 36 | 20 | 18225676 | 38 | 20 | 217437824 | 40 | 20 | 2187904502 |
| 36 | 21 | 6406616 | 38 | 21 | 88559862 | 40 | 21 | 1014394516 |
| 36 | 22 | 1615006 | 38 | 22 | 31817912 | 40 | 22 | 428257958 |
| 36 | 23 | 139760 | 38 | 23 | 9956348 | 40 | 23 | 159976584 |
| 36 | 24 | 1072 | 38 | 24 | 1733664 | 40 | 24 | 53453638 |
| | | | 38 | 25 | 81020 | 40 | 25 | 13638392 |
| | | | | | | 40 | 26 | 1582186 |
| | | | | | | 40 | 27 | 34972 |

Table 3: continued

| $n$ | $k$ | $p_n(k)$ |
|---|---|---|
| 42 | 0 | 61235363802 |
| 42 | 1 | 294873317972 |
| 42 | 2 | 762110650372 |
| 42 | 3 | 1399688241956 |
| 42 | 4 | 2055118460378 |
| 42 | 5 | 2568728314030 |
| 42 | 6 | 2838932155072 |
| 42 | 7 | 2843720976824 |
| 42 | 8 | 2626666404320 |
| 42 | 9 | 2265228537960 |
| 42 | 10 | 1840844618944 |
| 42 | 11 | 1419474224078 |
| 42 | 12 | 1043977291720 |
| 42 | 13 | 735151740056 |
| 42 | 14 | 496824634882 |
| 42 | 15 | 322875667652 |
| 42 | 16 | 201753866414 |
| 42 | 17 | 121230197502 |
| 42 | 18 | 69955721188 |
| 42 | 19 | 38635944222 |
| 42 | 20 | 20323389508 |
| 42 | 21 | 10146432880 |
| 42 | 22 | 4758325428 |
| 42 | 23 | 2059822102 |
| 42 | 24 | 805102310 |
| 42 | 25 | 279144480 |
| 42 | 26 | 83753436 |
| 42 | 27 | 16591468 |
| 42 | 28 | 1212792 |
| 42 | 29 | 10640 |

Table 3: continued

| $n$ | $k$ | $c_n(k)/4$ | $n$ | $k$ | $c_n(k)/4$ | $n$ | $k$ | $c_n(k)/4$ | $n$ | $k$ | $c_n(k)/4$ |
|---|---|---|---|---|---|---|---|---|---|---|---|
| 1 | 0 | 1 | 2 | 0 | 3 | 3 | 0 | 7 | 4 | 0 | 17 |
| | | | | | | 3 | 1 | 2 | 4 | 1 | 8 |
| 5 | 0 | 41 | 6 | 0 | 99 | 7 | 0 | 235 | 8 | 0 | 561 |
| 5 | 1 | 22 | 6 | 1 | 64 | 7 | 1 | 184 | 8 | 1 | 508 |
| 5 | 2 | 8 | 6 | 2 | 32 | 7 | 2 | 86 | 8 | 2 | 268 |
| | | | | | | 7 | 3 | 38 | 8 | 3 | 132 |
| | | | | | | | | | 8 | 4 | 10 |
| 9 | 0 | 1331 | 10 | 0 | 3167 | 11 | 0 | 7485 | 12 | 0 | 17753 |
| 9 | 1 | 1344 | 10 | 1 | 3556 | 11 | 1 | 9244 | 12 | 1 | 23876 |
| 9 | 2 | 850 | 10 | 2 | 2458 | 11 | 2 | 6900 | 12 | 2 | 19250 |
| 9 | 3 | 346 | 10 | 3 | 1152 | 11 | 3 | 3888 | 12 | 3 | 11436 |
| 9 | 4 | 196 | 10 | 4 | 596 | 11 | 4 | 1606 | 12 | 4 | 5660 |
| | | | | 5 | 96 | 11 | 5 | 888 | 12 | 5 | 2524 |
| | | | | | | 11 | 6 | 62 | 12 | 6 | 734 |
| 13 | 0 | 41867 | 14 | 0 | 99043 | 15 | 0 | 233157 | 16 | 0 | 550409 |
| 13 | 1 | 60884 | 14 | 1 | 154792 | 15 | 1 | 389792 | 16 | 1 | 979240 |
| 13 | 2 | 52934 | 14 | 2 | 143140 | 15 | 2 | 383628 | 16 | 2 | 1018166 |
| 13 | 3 | 33472 | 14 | 3 | 96904 | 15 | 3 | 276892 | 16 | 3 | 774040 |
| 13 | 4 | 19076 | 14 | 4 | 56594 | 15 | 4 | 169214 | 16 | 4 | 500926 |
| 13 | 5 | 7444 | 14 | 5 | 27300 | 15 | 5 | 91128 | 16 | 5 | 275232 |
| 13 | 6 | 3978 | 14 | 6 | 11310 | 15 | 6 | 37466 | 16 | 6 | 134610 |
| 13 | 7 | 720 | 14 | 7 | 4244 | 15 | 7 | 17324 | 16 | 7 | 53040 |
| | | | 14 | 8 | 284 | 15 | 8 | 5410 | 16 | 8 | 21890 |
| | | | | | | 15 | 9 | 138 | 16 | 9 | 3780 |
| 17 | 0 | 1293817 | 18 | 0 | 3048915 | | | | | | |
| 17 | 1 | 2442268 | 18 | 1 | 6080388 | | | | | | |
| 17 | 2 | 2681356 | 18 | 2 | 7008782 | | | | | | |
| 17 | 3 | 2149774 | 18 | 3 | 5894524 | | | | | | |
| 17 | 4 | 1459644 | 18 | 4 | 4168254 | | | | | | |
| 17 | 5 | 841890 | 18 | 5 | 2537728 | | | | | | |
| 17 | 6 | 444576 | 18 | 6 | 1362950 | | | | | | |
| 17 | 7 | 189650 | 18 | 7 | 658576 | | | | | | |
| 17 | 8 | 79632 | 18 | 8 | 267858 | | | | | | |
| 17 | 9 | 30716 | 18 | 9 | 105212 | | | | | | |
| 17 | 10 | 3346 | 18 | 10 | 30408 | | | | | | |
| | | | 18 | 11 | 1088 | | | | | | |

Table 4:

| $n$ | $k$ | $c_n(k)/4$ | $n$ | $k$ | $c_n(k)/4$ | $n$ | $k$ | $c_n(k)/4$ |
|---|---|---|---|---|---|---|---|---|
| 19 | 0 | 7158201 | 20 | 0 | 16843573 | 21 | 0 | 39504435 |
| 19 | 1 | 15049866 | 20 | 1 | 37200956 | 21 | 1 | 91512966 |
| 19 | 2 | 18207818 | 20 | 2 | 47034904 | 21 | 2 | 120863206 |
| 19 | 3 | 16046364 | 20 | 3 | 43256096 | 21 | 3 | 115919582 |
| 19 | 4 | 11829258 | 20 | 4 | 33149118 | 21 | 4 | 92235318 |
| 19 | 5 | 7530130 | 20 | 5 | 21896316 | 21 | 5 | 63319470 |
| 19 | 6 | 4240496 | 20 | 6 | 12912128 | 21 | 6 | 38842204 |
| 19 | 7 | 2170710 | 20 | 7 | 6763244 | 21 | 7 | 21312058 |
| 19 | 8 | 968778 | 20 | 8 | 3274210 | 21 | 8 | 10792706 |
| 19 | 9 | 387378 | 20 | 9 | 1369416 | 21 | 9 | 4893520 |
| 19 | 10 | 154960 | 20 | 10 | 518706 | 21 | 10 | 1986952 |
| 19 | 11 | 34190 | 20 | 11 | 183172 | 21 | 11 | 756634 |
| 19 | 12 | 1006 | 20 | 12 | 22452 | 21 | 12 | 239288 |
| | | | | | | 21 | 13 | 23168 |
| 22 | 0 | 92838503 | 23 | 0 | 217549387 | 24 | 0 | 510702499 |
| 22 | 1 | 224889896 | 23 | 1 | 550409212 | 24 | 1 | 1346063500 |
| 22 | 2 | 309216494 | 23 | 2 | 787511174 | 24 | 2 | 1998666370 |
| 22 | 3 | 308316464 | 23 | 3 | 815771144 | 24 | 3 | 2145565908 |
| 22 | 4 | 254062502 | 23 | 4 | 695840182 | 24 | 4 | 1890521138 |
| 22 | 5 | 180643016 | 23 | 5 | 511950948 | 24 | 5 | 1435323712 |
| 22 | 6 | 114457820 | 23 | 6 | 335858110 | 24 | 6 | 971353634 |
| 22 | 7 | 65452948 | 23 | 7 | 199169112 | 24 | 7 | 593724016 |
| 22 | 8 | 34023448 | 23 | 8 | 108005076 | 24 | 8 | 333701296 |
| 22 | 9 | 16342620 | 23 | 9 | 54030120 | 24 | 9 | 171821676 |
| 22 | 10 | 7029848 | 23 | 10 | 24789036 | 24 | 10 | 82408644 |
| 22 | 11 | 2663104 | 23 | 11 | 10292240 | 24 | 11 | 35940268 |
| 22 | 12 | 974308 | 23 | 12 | 3847090 | 24 | 12 | 14044418 |
| 22 | 13 | 219996 | 23 | 13 | 1358836 | 24 | 13 | 5056504 |
| 22 | 14 | 9154 | 23 | 14 | 263792 | 24 | 14 | 1523664 |
| | | | 23 | 15 | 7994 | 24 | 15 | 179920 |
| | | | | | | 24 | 16 | 2162 |
| 25 | 0 | 1195823247 | 26 | 0 | 2804575869 | 27 | 0 | 6562607385 |
| 25 | 1 | 3280337168 | 26 | 1 | 7989432672 | 27 | 1 | 19398952628 |
| 25 | 2 | 5052329956 | 26 | 2 | 12735745520 | 27 | 2 | 31990605456 |
| 25 | 3 | 5617273282 | 26 | 3 | 14635149660 | 27 | 3 | 37975439858 |

Table 4: Continued.

| $n$ | $k$ | $c_n(k)/4$ | $n$ | $k$ | $c_n(k)/4$ | $n$ | $k$ | $c_n(k)/4$ |
|---|---|---|---|---|---|---|---|---|
| 25 | 4 | 5110374048 | 26 | 4 | 13723257002 | 27 | 4 | 36687855574 |
| 25 | 5 | 4002750354 | 26 | 5 | 11066113256 | 27 | 5 | 30446746918 |
| 25 | 6 | 2793373142 | 26 | 6 | 7942250944 | 27 | 6 | 22474908138 |
| 25 | 7 | 1763705626 | 26 | 7 | 5160559796 | 27 | 7 | 15025151052 |
| 25 | 8 | 1023991992 | 26 | 8 | 3080396086 | 27 | 8 | 9237967176 |
| 25 | 9 | 548429998 | 26 | 9 | 1703470136 | 27 | 9 | 5264647012 |
| 25 | 10 | 273099178 | 26 | 10 | 873273892 | 27 | 10 | 2797593720 |
| 25 | 11 | 125831856 | 26 | 11 | 417566464 | 27 | 11 | 1387134186 |
| 25 | 12 | 53364032 | 26 | 12 | 184559436 | 27 | 12 | 642687898 |
| 25 | 13 | 20253608 | 26 | 13 | 74060220 | 27 | 13 | 276027632 |
| 25 | 14 | 7197122 | 26 | 14 | 26965768 | 27 | 14 | 107944142 |
| 25 | 15 | 2002618 | 26 | 15 | 8860896 | 27 | 15 | 38397784 |
| 25 | 16 | 201500 | 26 | 16 | 1853750 | 27 | 16 | 12289666 |
| 25 | | | 26 | 17 | 105188 | 27 | 17 | 2326206 |
| 25 | | | | | | 27 | 18 | 90476 |
| 28 | 0 | 15378643401 | 29 | 0 | 35964253315 | 30 | 0 | 84216378195 |
| 28 | 1 | 47081130896 | 29 | 1 | 113955364388 | 31 | 0 | 196843613381 |
| 28 | 2 | 80171819670 | 29 | 2 | 200285841574 | 32 | 0 | 460644961545 |
| 28 | 3 | 98143110328 | 29 | 3 | 252719748694 | | | |
| 28 | 4 | 97542797720 | 29 | 4 | 258304841600 | | | |
| 28 | 5 | 83171958968 | 29 | 5 | 226229739140 | | | |
| 28 | 6 | 63017743614 | 29 | 6 | 175916739126 | | | |
| 28 | 7 | 43234422648 | 29 | 7 | 123873896460 | | | |
| 28 | 8 | 27284382172 | 29 | 8 | 80239865186 | | | |
| 28 | 9 | 15960029252 | 29 | 9 | 48245510340 | | | |
| 28 | 10 | 8725745632 | 29 | 10 | 27125641348 | | | |
| 28 | 11 | 4452294048 | 29 | 11 | 14304265858 | | | |
| 28 | 12 | 2130127362 | 29 | 12 | 7085279050 | | | |
| 28 | 13 | 950315284 | 29 | 13 | 3293828278 | | | |
| 28 | 14 | 389858002 | 29 | 14 | 1432023556 | | | |
| 28 | 15 | 145741920 | 29 | 15 | 573273466 | | | |
| 28 | 16 | 49190166 | 29 | 16 | 209372954 | | | |
| 28 | 17 | 13502664 | 29 | 17 | 69353438 | | | |
| 28 | 18 | 1799226 | 29 | 18 | 18180612 | | | |
| 28 | 19 | 32588 | 29 | 19 | 2011954 | | | |
| | | | 29 | 20 | 26996 | | | |

Table 4: Continued.